%% file: two.tex
\documentstyle[twocolumn,aps]{revtex}
\input psbox
\begin{document}
\title{
Two-Color Coherent Photodissociation of Nitrogen Oxide in Intense Laser
Fields}
\author{Julian Juhi-Lian Ting}

\address{Physics Institute, Ching-hua University, \\
Hsin-Chu, Taiwan  30043, Republic of China \footnote{E-mail address:
%\address{Department of Physics, Tsing-hua University, \\
%Hsinchu, Taiwan  30043, Republic of China \footnote{E-mail address:
jlting@phys.nthu.edu.tw \\
~Present address: \\
~Institute f\"ur Festk\"orperforschung,\\
~Institut 5 -  Neutronenstreung,\\
~Forschungszentrum J\"ulich GmbH\\
~Postfach 1913\\
~D-52428 J\"ulich, Germany
}}

\date{\today}
\maketitle
\begin{abstract}
A simple one-dimensional semi-classical model with a Morse potential
is used to investigate the possibility of two-color infrared multi-photon 
dissociation of vibrationally excited nitrogen oxide.
The amplitude ratio effects and adiabatic effects are investigated.
Some initial states are found to have thresholds smaller than
expected from single-mode considerations and multiple thresholds exist
for initial states up to 32.

\end{abstract}
%\section{Introduction}
Two-color
lasers interacting with molecules with a phase difference might alter
greatly
the behavior of molecular dissociation. 
%Two-mode lasing with a fixed phase difference can be practically
%achieved through various laser cavity configurations.
%There are various configurations of lasers that produces two-mode lasing 
%with a fixed phase difference.
%For some of the pump-probe laser-matter interaction
%system the phase difference between two
%lasers are important while others not, like a stimulated Raman process
%involving adiabatic passage (STIRAP) method\cite{GRSB} and  frequency
%chirping method\cite{MGHTW}. 
%In this paper we
%will focus on the former case. There are also laser cavities generate
%muti-mode lasers and for mode-lock lasers the phases between modes are
%fixed, which will have the same effects as two lasers with a fixed phase
%difference.
%Some of them use one laser cavity while others use
%several laser cavities.
According to the calculation by Charron {\it et al} for $H_2^+$ with varied 
phase shifts between two laser fields,\cite{CGM} the above-threshold 
dissociation of $H_2^+$ can be strongly enhanced by combining the laser
radiation with a harmonic and varying the relative phase
between the two radiations.
However their calculation is for only the first excited state.

It has become possible to transfer population of nitrogen oxide (NO) 
efficiently to higher states:
Yang {\it et al }
reported experiments in which 
molecules were populated into an initial vibrational state as great as $
n = 25 $
by stimulated-emission pumping (SEP)\cite{YKW}
 whereas Schiemann {\it et al }
showed 
how to populate to a state $n = 6$ by a stimulated
Raman process involving adiabatic passage (STIRAP) using pulsed
lasers\cite{SKSB}.
Therefore it is desirable to know about the behavior of a diatomic
molecular at other excited states
under intense two-color laser fields.

%In a previous paper Morse potential \cite{M} was used 
%to calculate the dissociation of NO
%under single mode lasers.\cite{T}
In this work, previous results \cite{T} for single-mode lasers using a
Morse potential were extended
for a laser with its own
third harmonic.
Instead of considering $H_2^+$, as did by
Charron {\it et al}, we considered NO, which 
can serve to answer the question whether the
observed behavior is universal for all diatomic molecules. 
The main purpose of this work was to find the dissociation thresholds of
NO with two-color lasers; guided  by results of two-color
lasers we returned to find lower thresholds under single-mode.

The model used is simple and cannot distinguish between above-threshold
dissociation and other effects such as stimulated emission, 
multiphoton absorption, 
bond softening  and suppression of dissociation 
%due to temporary
%vibrational trapping in the upper potential wells 
as did Charron {\it et al}.
However, we can assess which effect has the most significant influence.
As above-threshold
dissociation is the most important contribution for photodissociation of
molecules, 
according to Charron 
{\it et al},  
one would expect our results to be consistent with theirs.
%The lasers calculated is ?? pico-second pulsed laser at a basic
%wave length ??? nm.

%\section{Model}                            
We consider an isolated non-rotating NO molecule interacting
with plane-polarized harmonic laser fields.
The dimensionless Hamiltonian of a free diatomic molecule as a Morse
oscillator is
\begin{equation}
H_0={p^2 \over 2} + {(1-e^{- z})^2 \over 2},
\end{equation}
in which
$z$ and $p$  are the dimensionless coordinate and the corresponding momentum
respectively.
The parameters of NO are given 
in Ref. \cite{T}. 
%by Huber and Herzberg\cite{HH}  and summarized
%in table \ref{mdata}.               

The effects of neglecting the rotational contribution are two fold.
Firstly, additional selection rules might apply if one considered the
rotational contribution. 
Secondly, 
angular momentum contributes an effective
potential $l (l+1) / (z+ r_e)^2$, with $r_e$ the equilibrium inter-nuclei
distance,  which alters the asymptotic behavior
of the effective potential of the radial Schr\"odinger  equation
at large separation.\cite{DN}
%Without the angular part the potential
%approaches infinity from below unity, with the angular part it
%approaches from above.
The Hamiltonian without the angular part corresponds to 
a model of DNA denaturation in the continuum
limit.\cite{DPB}

We concluded in previous work that an exponential form of
coupling is suitable for molecular calculation. 
With two color-lasers the interacting Hamiltonian reads
\begin{equation}
{{A_0 \Omega} \over 2}{(z+a) e^{-(z+a)/b}} (A_1 \sin (\Omega t) +
A_2 \sin ( 3 \Omega t + \phi))
\sin^2
({{2 \pi t} \over T}),
\label{MH}
\end{equation}
in which $\Omega$ and $A_j$, $j = 0, 1, 2 $, are dimensionless frequency and amplitudes
respectively.
%, and $q$ is the effective dipole charge of the molecule.
%Many previous authors\cite{HM2,HM1,TM2,TM1} have used coupling functions of
%similar form.
The variables $a$ and $b$ were chosen to have the values 2 and 1
respectively in the calculation.
The definition of the dissociation rate is
\cite{HM2}
\begin{equation}
 P_{diss} (t) = 1 - < \psi_I (t)  | \psi_I (t) >,
\end{equation}
%This definition is appropriate if coherent excitation become
%important.
%\cite{SS}
in which $| \psi_I (t) >$ is the interacting wavepacket.
We take $T$, the pulse duration, to be 300 (or 600) optical cycles.
%This choice has the
%advantage that, when $A=A_c$, $T_d$ is almost 300 (or 600) cycles and the field
%strength is almost zero at that time. 
With this choice the pulse
duration lies in the nanosecond range. 
The dimensionless amplitude $A = 1$
corresponds to a laser intensity $320~TW / cm^2$.

%From the point of view of phase transition, if we varying the amplitude
%of these two sinusoidal wave form to see the change of the dissociation
%threshold it is a problem consisting
%competition between two length scale, therefore a commensurable and
%incommensurable phase transition may  occur. However, in our numerical
%calculation we can only probe commensurate phase because numerical methods
%have no way to represent irrational numbers. From another point
%of view we are simply varying the driving wave form to a more complex
%one. 

%\section{Numerical results}

%The following numerical results were obtained using a method
%described and verified
%previously by Ting {\it et al} \cite{TYJ}, which 
The numerical method used is a fast Fourier-transformed grid
method. \cite{TYJ}
%also considered by other authors
%\cite{F,ALL}.
All calculations were done dimensionlessly at $\Omega = 0.9$, which is
slightly less than the transition frequency between the ground state and
the first excited state of the free Morse oscillator.
Each dissociation threshold curve was obtained from more than 1000
points of calculations.
There is no general rule to find out a second or a third transition for
each initial state. However, it is reasonable to expect that neighboring
states, or
other similar states,
should have similar thresholds. 

%\subsection {Single Mode Lasers}

Firstly, the previous result of a single-mode laser is reproduced in
fig.1, with
several additional lower thresholds found and $n_0 = 1$ is found
to have three thresholds.
%The values are tabulated in table {\ref {dfig1}}

%\subsection {Two-color Lasers}
Secondly, for two-color lasers, four curves 
for $A_1 = 1 $ and $A_2 = \sqrt 3$ with  $\phi
= 0, 
{\pi \over 2},
\pi,
{{3 \pi} \over 2}$ in Eq. (\ref{MH})
are plotted in fig.2. 
As before we computed both $T=300$ and $T=600$.
%, and the corresponding data were listed in table
%\ref{dfig2a}, \ref{dfig2b}, \ref{dfig2c}, \ref{dfig2d}.
For over half the initial states
a phase difference $\pi$ reduce the value of $A_c$ most effectively.
In order to compare the effect of the phase difference, we replotted fig.2
in fig.3 for $T = 300$ and $600$ respectively. 
Although we simulated with more than twice the laser power in the present case
than in the single-mode case, the threshold amplitudes were not diminished
correspondingly, contrary to results of Charron {\it et al}.
%except for some states like ...... become
%significantly lower with some particular combination of phase difference
%and amplitudes. 
By comparing $A_c$ for $T=300$ and for $T=600$ one obtains 
hints about whether
a particular value of $n_0$ is dissociated by itself or by transition to other
states. According to perturbation theories\cite{RS} 
the shorter is the light pulse, the more stationary states are
excited and {\it vice versa}. 
%And a very long laser pulse excites only
%that particular stationary state in the upper manifold which is in
%resonance with the electromagnetic field.
The fact that a phase difference $\pi $  does not invariably result in the
smallest dissociation threshold
contrasts with our intuition that zero phase difference 
has the largest intensity, 
but is
consistent with the calculation on first excited states by Charron {\it et al}.
However, the difference
diminished as $T$ became larger, i.e. fig. 3(b). Multiple thresholds
were found up to an initial state equal to 32 for a phase difference 
$\pi$.

Thirdly, a modified amplitude ratio for $A_1 = 1$ and $A_2 = 2$ 
was also investigated;
the results are plotted
in fig.4 with $T = 300$, comparison with fig.3(a). 
At a phase difference $\pi / 2$,
$n_0 = 4$
has four thresholds. A phase difference $\pi$  still has the most lowest
thresholds whereas a phase difference $\pi / 2 $ has no lowest threshold.

There are some inherent limitations to this model.
One suspects that multiple thresholds at higher initial states
can also be found with calculations of finer amplitude resolution.
Secondly, $n_0 > 44$ shows almost the same $A_c$ in all calculations,
which is also attributed to the resolution of $A$ calculated. 
However, consideration of spectral data \cite{HH} shows that
%start from the ground state and 
above the vibrational state $v=30$ an electronically 
excited state $ ^2 \Sigma^+$ can be excited
by a radiationless transition from the ground electronic state.
Furthermore NO has an ionization energy $9.26~eV$ somewhat greater than the
dissociation energy, $6.5~eV$. Therefore under strong fields the molecule might be excited electronically and
suffer ionization before dissociation.

%\section{Summary}
%In summary, 
%thresholds of vibrational dissociation of NO under two-color lasers are 
%considered in this paper.  The effects
%of adiabatic switching and the effects of amplitude ratios were
%considered. 
In summary, the following conclusions are found in comparison with single-mode 
driven lasers in previous work.
Firstly, up to four  thresholds were found for two-color
laser cases instead of three thresholds for single-mode laser excitation.
Secondly, multiple transitions were found for initial states as high as
32, instead of 20 under a single-mode laser.
Both the first and second points support that two-color cases are
more complicated than single-mode cases.
Thirdly, the tail of the dissociation curve remains almost constant,
perhaps because our
computation cannot resolve that fine detail.
Fourthly, for most initial states the power required is not much 
influenced by the presence of
the higher harmonic laser, although some initial states
with smaller dissociation threshold were found. This conclusion 
disagrees with
results of Charron {\it et al}.
Finally, our result is consistent with those of Charron {\it et al}
that for the first excited state a phase difference $\phi = \pi$ 
results in the greatest dissociation rate, although rotational
effects are not yet considered.
Because of
the computational cost, only limited data were calculated.

%\section{Acknowledgments}
I thank Professor How-Sen Wong for his constant support and
Dr. J.F. Ogilvie for discussions.

%\newpage
\begin{figure}[p]
\begin{center}
\mbox{\psboxto(5cm;5cm){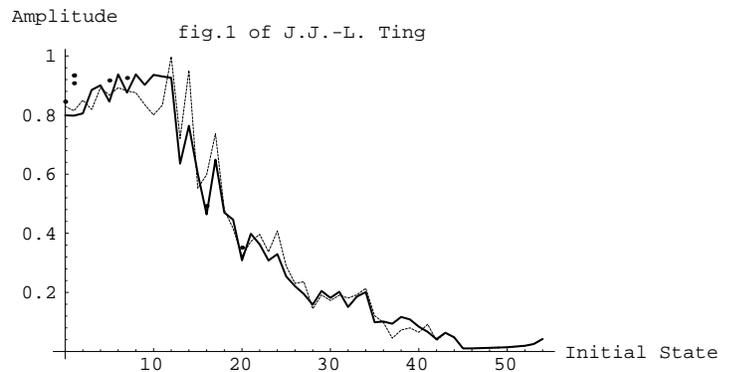}}
\caption[1]{Critical amplitude ($A_c$) of various initial states 
with a single-mode laser at $\Omega
=0.9$; for the solid thick line $T=300$ whereas for the thin dashed line
$T=600$.}
\end{center}
\end{figure}

\begin{figure}[bth]
\begin{center}
\mbox{\psboxto(8.5cm;5cm){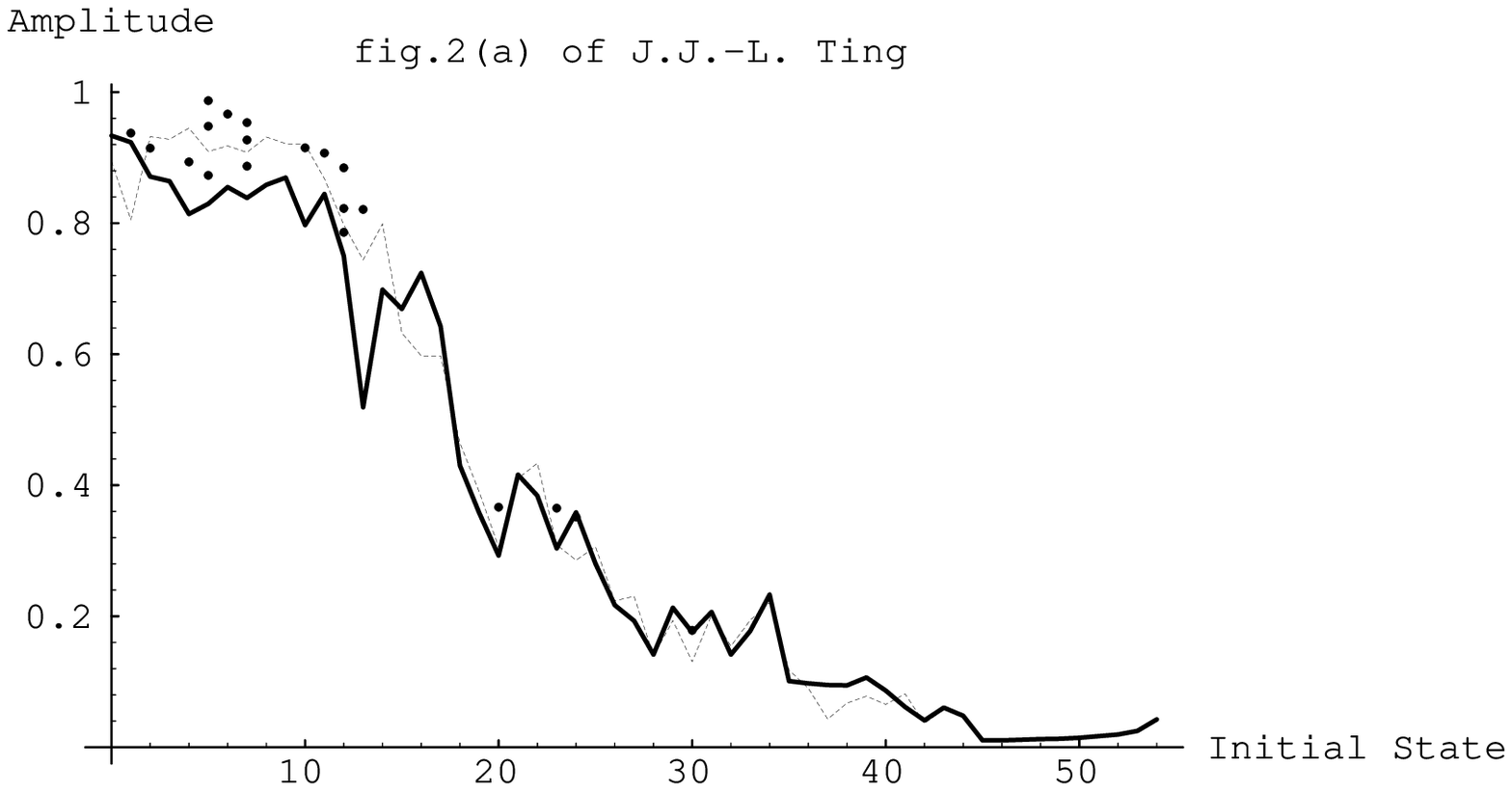}}
\mbox{\psboxto(8.5cm;5cm){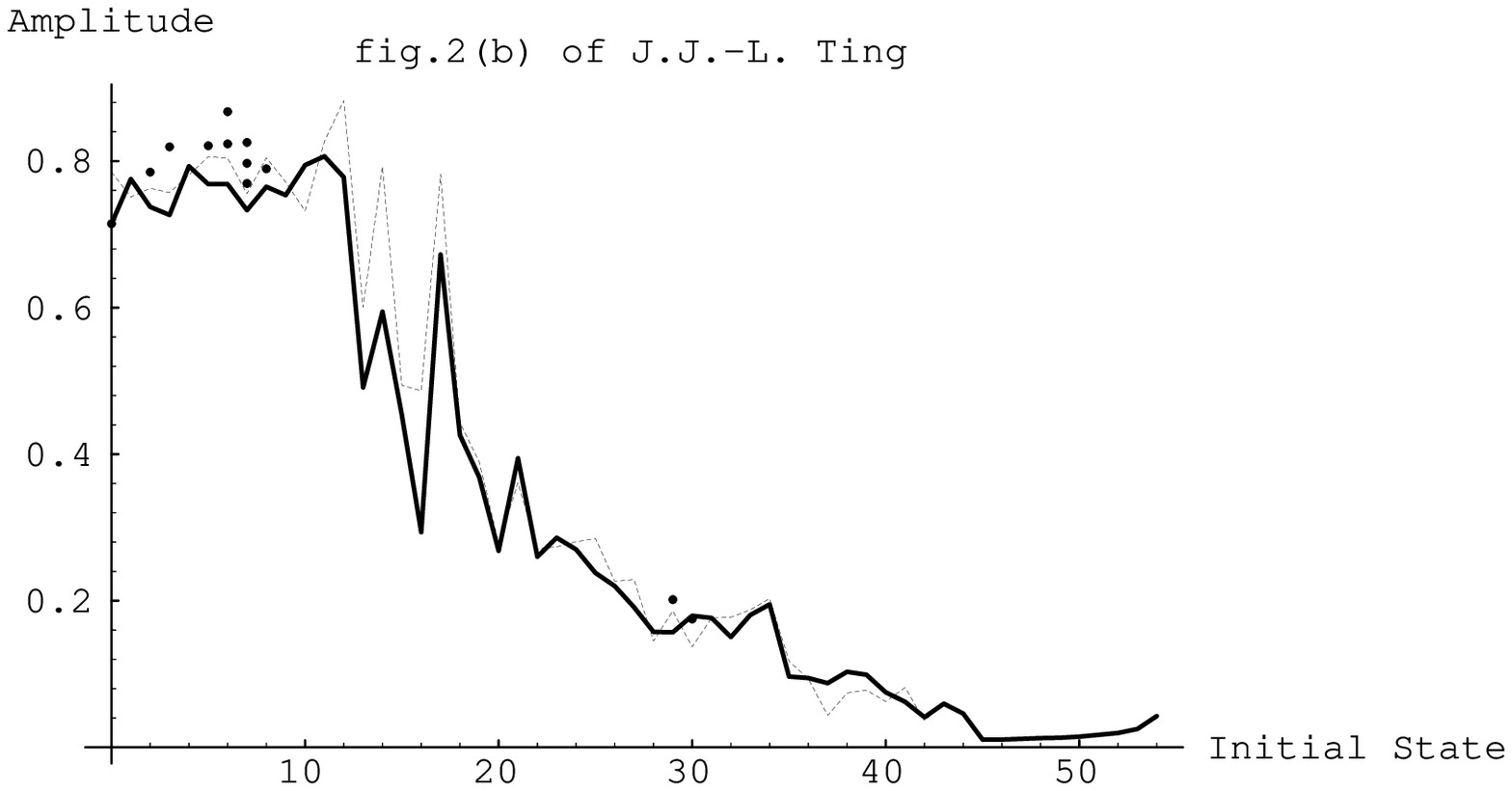}}
\mbox{\psboxto(8.5cm;5cm){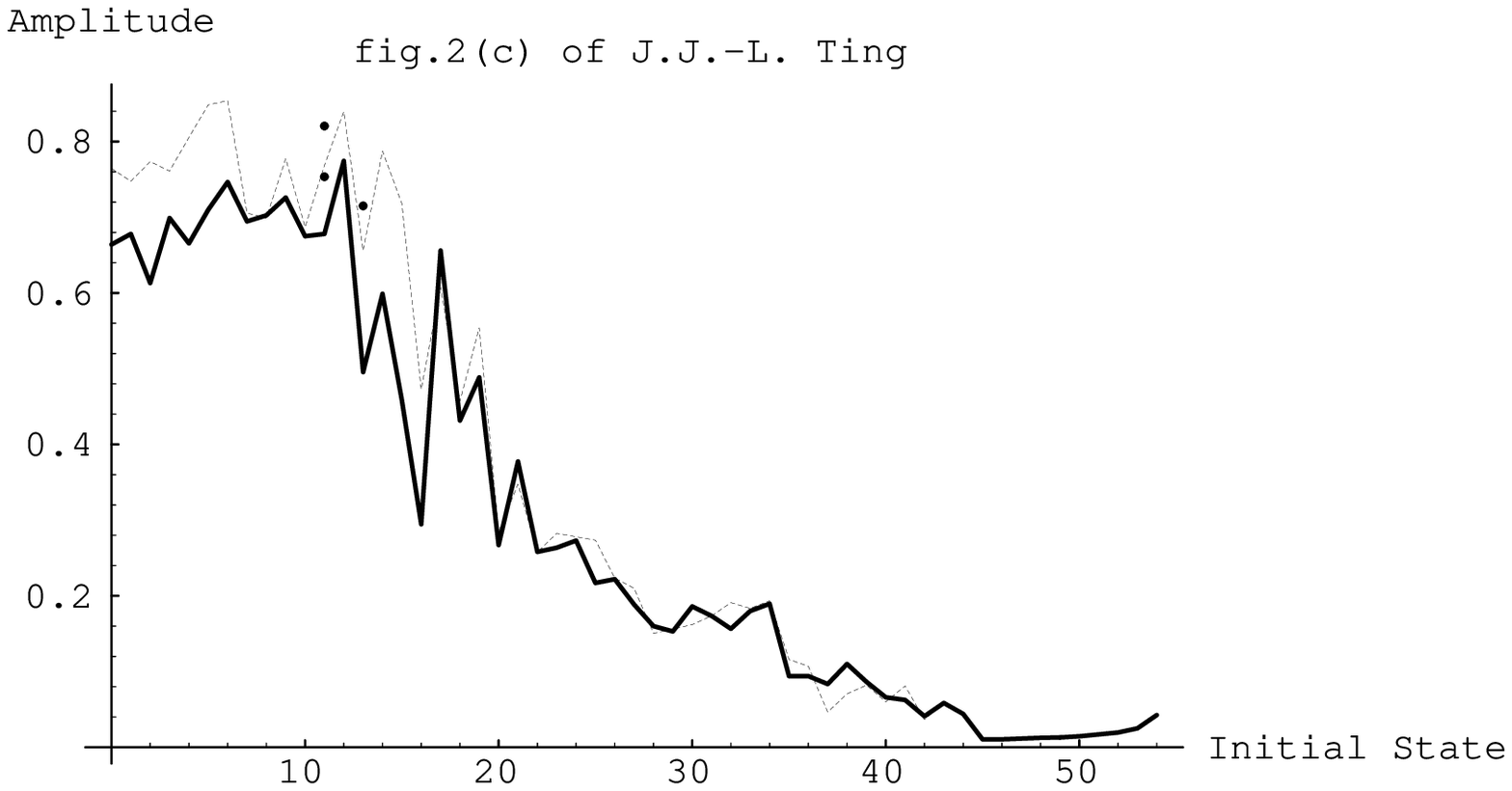}}
\mbox{\psboxto(8.5cm;5cm){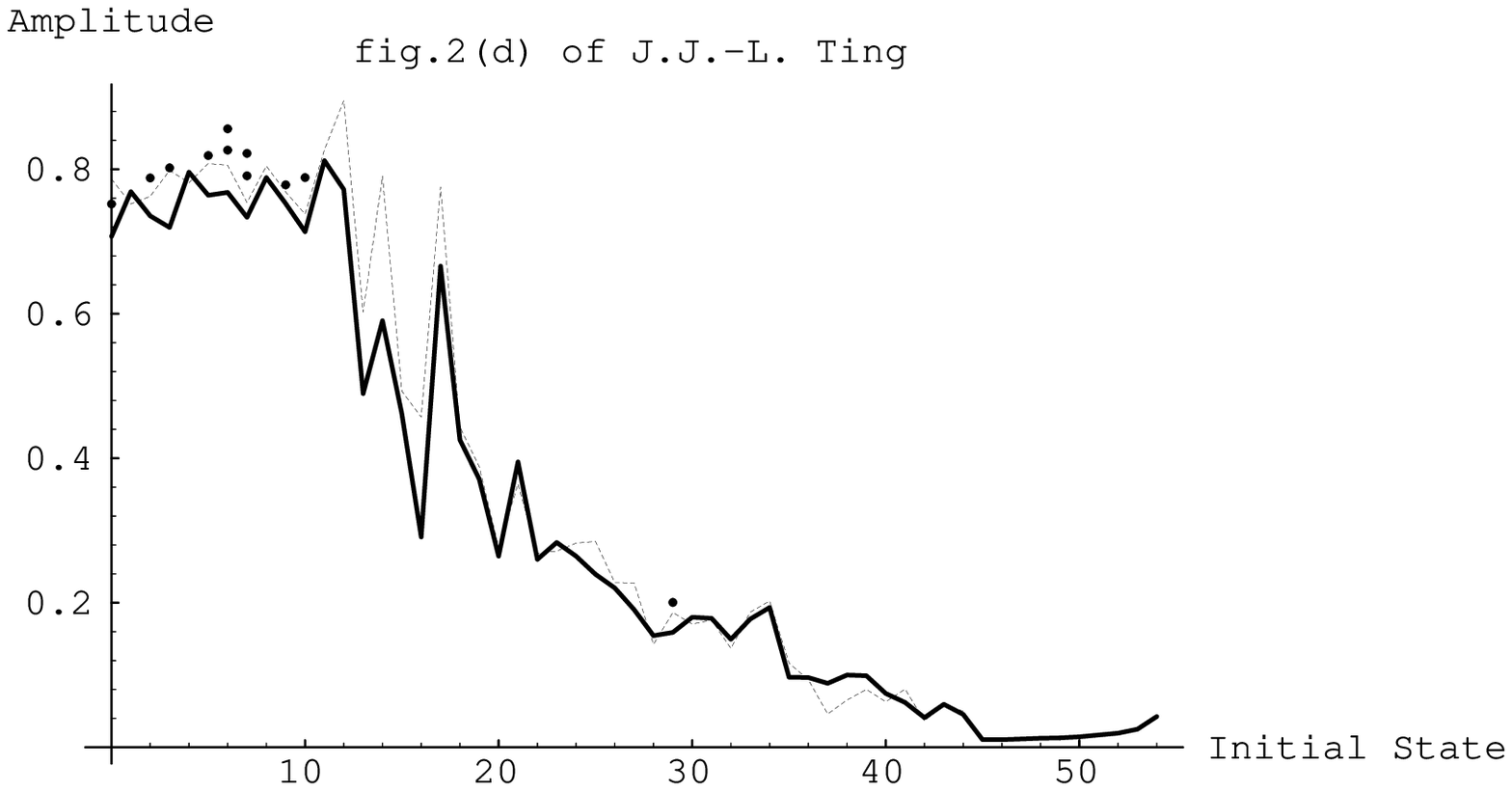}}
\caption[2]{Critical amplitudes ($A_c$) of various initial states with two
lasers at $A_1 = 1$, $A_2 = \sqrt 3 $ and $\Omega = 0.9$ in 
Eq. (\ref{MH}) at phase difference $\phi = $  
(a) $0$,
(b) ${\pi \over 2}$, 
(c) $\pi$,
(d) ${{3 \pi} \over 2}$ with  $T = 300$ (solid line) or  
$T= 600$ (dashed line).}
\end{center}
\end{figure}

\begin{figure}[bth]
\begin{center}
\mbox{\psboxto(8.5cm;5cm){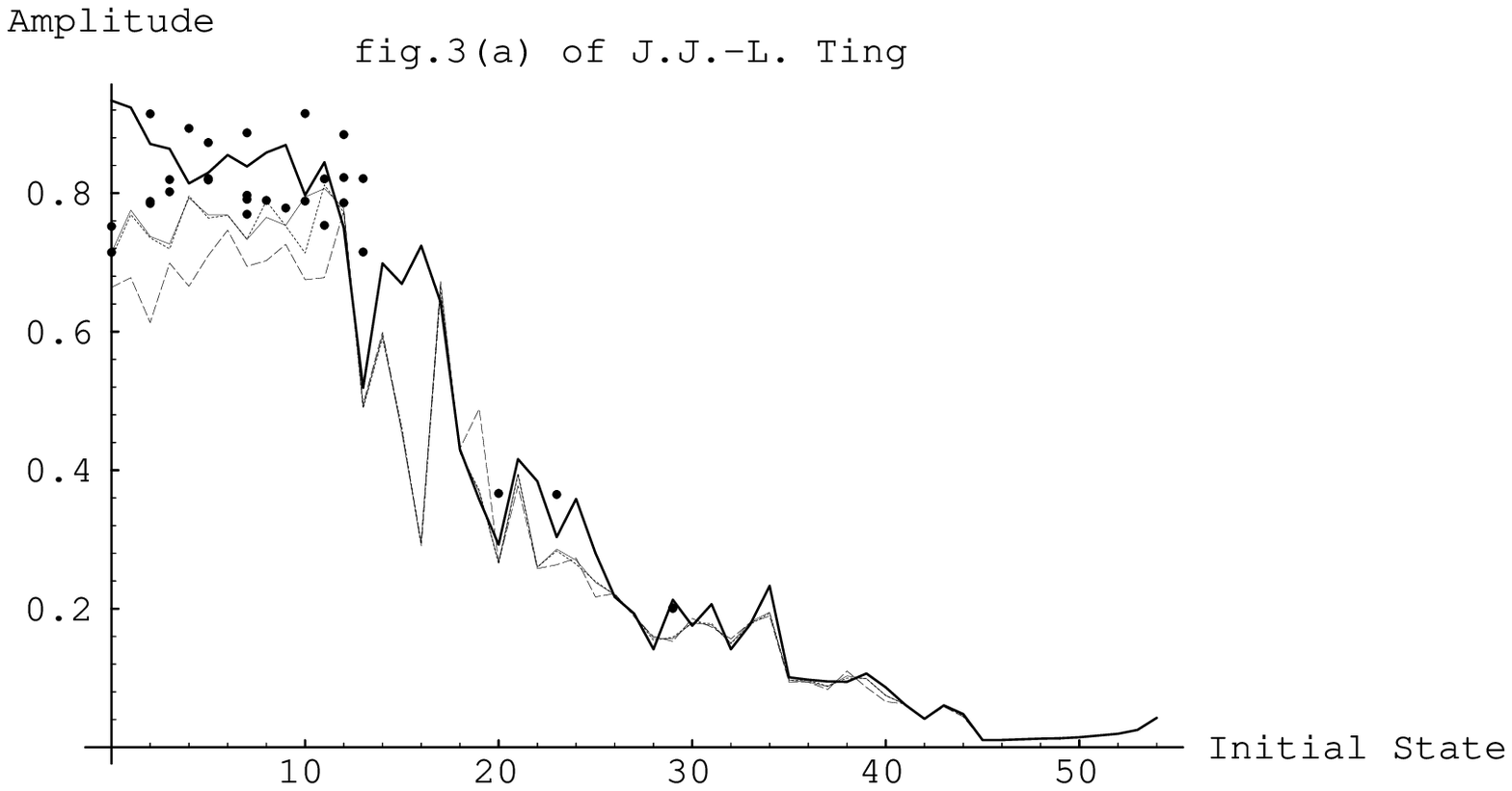}}
\mbox{\psboxto(8.5cm;5cm){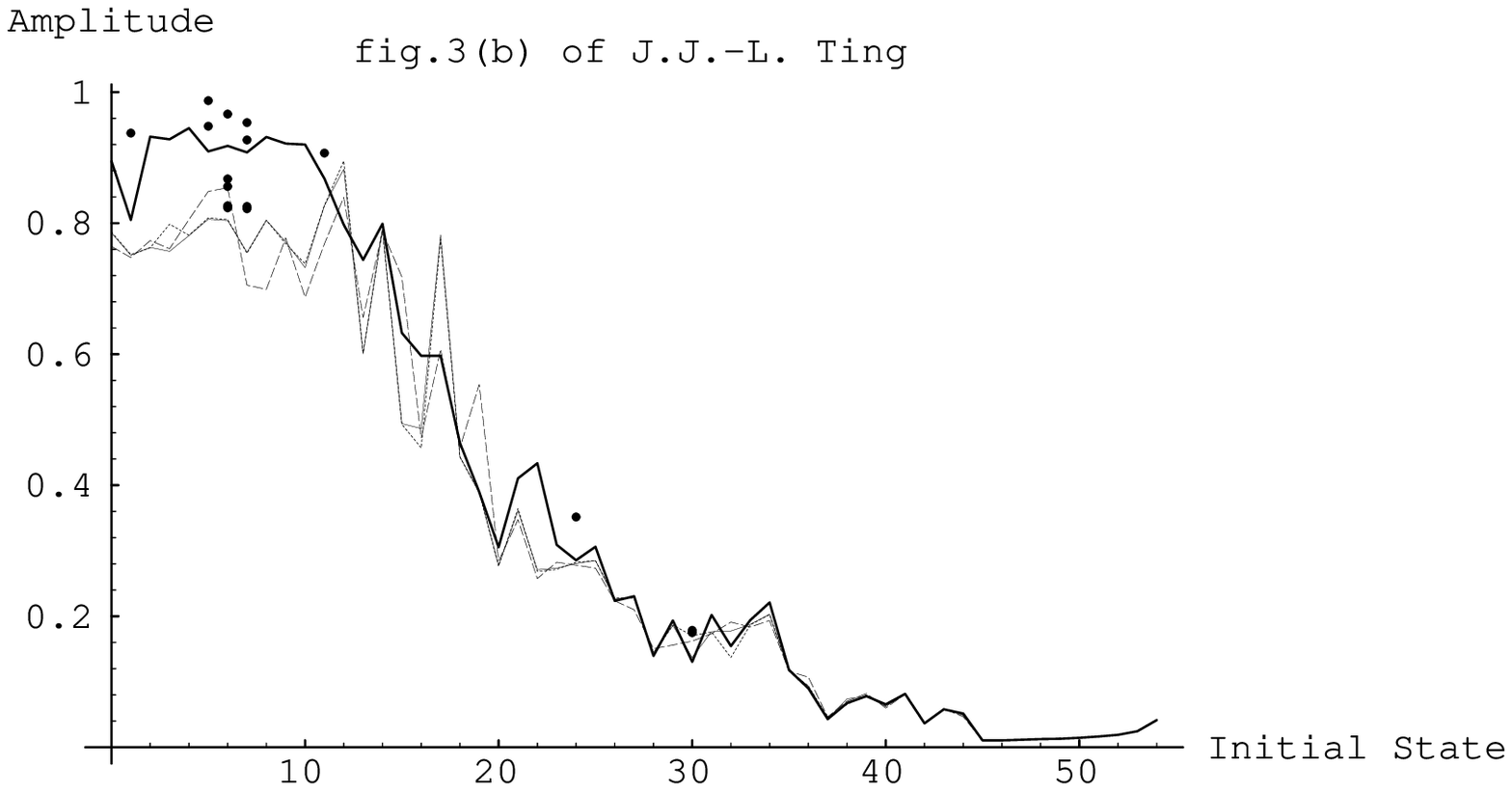}}
\caption[3]{Critical amplitudes ($A_c$) of various initial states for two
lasers at $A_1 = 1$, $A_2 = \sqrt 3 $ and $\Omega = 0.9$ in Eq. (\ref{MH})
at phase difference $\phi = $ 
$ 0$ (thick solid lines),
${\pi \over 2}$ (thin solid lines), 
$\pi$ (thick long dashed lines), and
${{3 \pi} \over 2}$ (thin short dashed lines)
with envelope periods (a) $T = 300$ or  (b) $T = 600$.}
\end{center}
\end{figure}

\begin{figure}[bth]
\begin{center}
\mbox{\psboxto(5cm;5cm){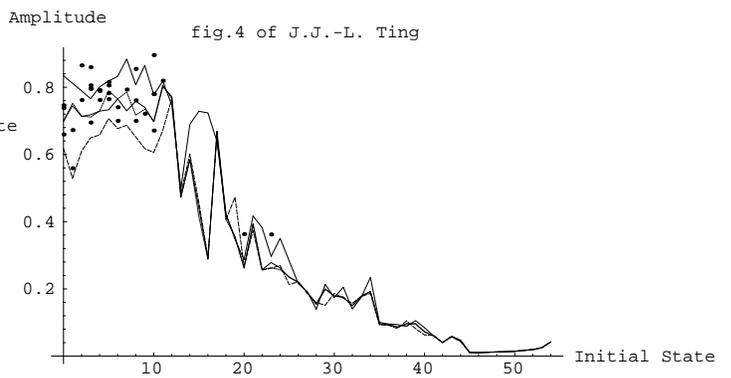}}
\caption[4]{Critical amplitude ($A_c$) of various initial states by two
lasers at $A_1 = 1$ and $A_2 = 2 $ in Eq. (\ref{MH})
at phase difference $\phi = $ 
$ 0$ (thick solid lines),
${\pi \over 2}$ (thin solid lines), 
$\pi$ (thick long dashed lines), and
${{3 \pi} \over 2}$ (thin short dashed lines)
with envelope period $T = 300$.}
\end{center}
\end{figure}

\end{document}

%% file: psbox.tex
\def\temp{1.34}%
\let\tempp=\relax
\expandafter\ifx\csname psboxversion\endcsname\relax
\else
    \ifdim\temp cm>\psboxversion cm
    \else
      \let\temp=\psboxversion
      \let\tempp= 
    \fi
\fi
\tempp
\let\psboxversion=\temp
\catcode`\@=11
\def\psfortextures{%     For TeXtures on the Macintosh
\def\PSspeci@l##1##2{%
\special{illustration ##1\space scaled ##2}%
}}%
\def\psfordvitops{%      For the DVItoPS converter on IBM mainframes
\def\PSspeci@l##1##2{%
\special{dvitops: import ##1\space \the\drawingwd \the\drawinght}%
}}%
\def\psfordvips{%      For DVIPS converter on VAX, UNIX and PC's
\def\PSspeci@l##1##2{%
\d@my=0.1bp \d@mx=\drawingwd \divide\d@mx by\d@my% BUG! for large \drawingwd
\includegraphics{##1\space}}}%
\def\psforoztex{%        For the OzTeX shareware on the Macintosh
\def\PSspeci@l##1##2{%
\special{##1 \space
      ##2 1000 div dup scale
      \number-\psllx\space \number-\pslly\space translate
}}}%
\def\psfordvitps{%       From the UNIX TeXPS package, vers.>3.12
\def\psdimt@n@sp##1{\d@mx=##1\relax\edef\psn@sp{\number\d@mx}}
\def\PSspeci@l##1##2{%
\special{dvitps: Include0 "psfig.psr"}% contains def of "startTexFig"
\psdimt@n@sp{\drawingwd}
\special{dvitps: Literal "\psn@sp\space"}
\psdimt@n@sp{\drawinght}
\special{dvitps: Literal "\psn@sp\space"}
\psdimt@n@sp{\psllx bp}
\special{dvitps: Literal "\psn@sp\space"}
\psdimt@n@sp{\pslly bp}
\special{dvitps: Literal "\psn@sp\space"}
\psdimt@n@sp{\psurx bp}
\special{dvitps: Literal "\psn@sp\space"}
\psdimt@n@sp{\psury bp}
\special{dvitps: Literal "\psn@sp\space startTexFig\space"}
\special{dvitps: Include1 "##1"}
\special{dvitps: Literal "endTexFig\space"}
}}%
\def\psfordvialw{%   Try for dvialw, a UNIX public domain
\def\PSspeci@l##1##2{
\special{language "PostScript",
position = "bottom left",
literal "  \psllx\space \pslly\space translate
  ##2 1000 div dup scale
  -\psllx\space -\pslly\space translate",
include "##1"}
}}%
\def\psforptips{%   For MS-DOS; LUOMA@brandeis.bitnet
\def\PSspeci@l##1##2{{
\d@mx=\psurx bp
\advance \d@mx by -\psllx bp
\divide \d@mx by 1000\multiply\d@mx by \xscale
\incm{\d@mx}
\let\tmpx\dimincm
\d@my=\psury bp
\advance \d@my by -\pslly bp
\divide \d@my by 1000\multiply\d@my by \xscale
\incm{\d@my}
\let\tmpy\dimincm
\d@mx=-\psllx bp
\divide \d@mx by 1000\multiply\d@mx by \xscale
\d@my=-\pslly bp
\divide \d@my by 1000\multiply\d@my by \xscale
\at(\d@mx;\d@my){\special{ps:##1 x=\tmpx, y=\tmpy}}
}}}%
\def\psonlyboxes{%     Draft-like behaviour if none of the others works
\def\PSspeci@l##1##2{%
\at(0cm;0cm){\boxit{\vbox to\drawinght
  {\vss\hbox to\drawingwd{\at(0cm;0cm){\hbox{({\tt##1})}}\hss}}}}
}}%
\def\psloc@lerr#1{%
\let\savedPSspeci@l=\PSspeci@l%
\def\PSspeci@l##1##2{%
\at(0cm;0cm){\boxit{\vbox to\drawinght
  {\vss\hbox to\drawingwd{\at(0cm;0cm){\hbox{({\tt##1}) #1}}\hss}}}}
\let\PSspeci@l=\savedPSspeci@l% restore normal output for other figs!
}}%
\newread\pst@mpin
\newdimen\drawinght\newdimen\drawingwd
\newdimen\psxoffset\newdimen\psyoffset
\newbox\drawingBox
\newcount\xscale \newcount\yscale \newdimen\pscm\pscm=1cm
\newdimen\d@mx \newdimen\d@my
\newdimen\pswdincr \newdimen\pshtincr
\let\ps@nnotation=\relax
{\catcode`\|=0 |catcode`|\=12 |catcode`|%=12 |catcode`~=12
|catcode`#=12 |catcode`*=14
|xdef|backslashother{\}*
|xdef|percentother{%}*
|xdef|tildeother{~}*
|xdef|sharpother{#}*
}%
\def\R@moveMeaningHeader#1:->{}%
\def\uncatcode#1{%
\edef#1{\expandafter\R@moveMeaningHeader\meaning#1}}%
\def\execute#1{#1}% NOT stupid: cs in #1 are then identified BEFORE execution
\def\psm@keother#1{\catcode`#112\relax}% borrowed from latex
\def\executeinspecs#1{%
\execute{\begingroup\let\do\psm@keother\dospecials\catcode`\^^M=9#1\endgroup}}%
\def\@mpty{}%
\def\matchexpin#1#2{
  \fi%
  \edef\tmpb{{#2}}%
  \expandafter\makem@tchtmp\tmpb%
  \edef\tmpa{#1}\edef\tmpb{#2}%
  \expandafter\expandafter\expandafter\m@tchtmp\expandafter\tmpa\tmpb\endm@tch%
  \if\match%
}%
\def\matchin#1#2{%
  \fi%
  \makem@tchtmp{#2}%
  \m@tchtmp#1#2\endm@tch%
  \if\match%
}%
\def\makem@tchtmp#1{\def\m@tchtmp##1#1##2\endm@tch{%
  \def\tmpa{##1}\def\tmpb{##2}\let\m@tchtmp=\relax%
  \ifx\tmpb\@mpty\def\match{YN}%
  \else\def\match{YY}\fi%
}}%
\def\incm#1{{\psxoffset=1cm\d@my=#1
 \d@mx=\d@my
  \divide\d@mx by \psxoffset
  \xdef\dimincm{\number\d@mx.}
  \advance\d@my by -\number\d@mx cm
  \multiply\d@my by 100
 \d@mx=\d@my
  \divide\d@mx by \psxoffset
  \edef\dimincm{\dimincm\number\d@mx}
  \advance\d@my by -\number\d@mx cm
  \multiply\d@my by 100
 \d@mx=\d@my
  \divide\d@mx by \psxoffset
  \xdef\dimincm{\dimincm\number\d@mx}
}}%
\newif\ifNotB@undingBox
\newhelp\PShelp{Proceed: you'll have a 5cm square blank box instead of
your graphics (Jean Orloff).}%
\def\s@tsize#1 #2 #3 #4\@ndsize{
  \def\psllx{#1}\def\pslly{#2}%
  \def\psurx{#3}\def\psury{#4}%  needed by a crazyness of dvips!
  \ifx\psurx\@mpty\NotB@undingBoxtrue% this is not a valid one!
  \else
    \drawinght=#4bp\advance\drawinght by-#2bp
    \drawingwd=#3bp\advance\drawingwd by-#1bp
  \fi
  }%
\def\sc@nBBline#1:#2\@ndBBline{\edef\p@rameter{#1}\edef\v@lue{#2}}%
\def\g@bblefirstblank#1#2:{\ifx#1 \else#1\fi#2}%
{\catcode`\%=12
\xdef\B@undingBox{%%BoundingBox}}%
\def\ReadPSize#1{
 \readfilename#1\relax
 \let\PSfilename=\lastreadfilename
 \openin\pst@mpin=#1\relax
 \ifeof\pst@mpin \errhelp=\PShelp
   \errmessage{I haven't found your postscript file (\PSfilename)}%
   \psloc@lerr{was not found}%
   \s@tsize 0 0 142 142\@ndsize
   \closein\pst@mpin
 \else
   \if\matchexpin{\GlobalInputList}{, \lastreadfilename}%
   \else\xdef\GlobalInputList{\GlobalInputList, \lastreadfilename}%
     \immediate\write\psbj@inaux{\lastreadfilename,}%
   \fi%
   \loop
     \executeinspecs{\catcode`\ =10\global\read\pst@mpin to\n@xtline}%
     \ifeof\pst@mpin
       \errhelp=\PShelp
       \errmessage{(\PSfilename) is not an Encapsulated PostScript File:
           I could not find any \B@undingBox: line.}%
       \edef\v@lue{0 0 142 142:}%
       \psloc@lerr{is not an EPSFile}%
       \NotB@undingBoxfalse
     \else
       \expandafter\sc@nBBline\n@xtline:\@ndBBline
       \ifx\p@rameter\B@undingBox\NotB@undingBoxfalse
         \edef\t@mp{%
           \expandafter\g@bblefirstblank\v@lue\space\space\space}%
         \expandafter\s@tsize\t@mp\@ndsize
       \else\NotB@undingBoxtrue
       \fi
     \fi
   \ifNotB@undingBox\repeat
   \closein\pst@mpin
 \fi
\message{#1}%
}%
\def\psboxto(#1;#2)#3{\vbox{%
   \ReadPSize{#3}%
   \advance\pswdincr by \drawingwd
   \advance\pshtincr by \drawinght
   \divide\pswdincr by 1000
   \divide\pshtincr by 1000
   \d@mx=#1
   \ifdim\d@mx=0pt\xscale=1000
         \else \xscale=\d@mx \divide \xscale by \pswdincr\fi
   \d@my=#2
   \ifdim\d@my=0pt\yscale=1000
         \else \yscale=\d@my \divide \yscale by \pshtincr\fi
   \ifnum\yscale=1000
         \else\ifnum\xscale=1000\xscale=\yscale
                    \else\ifnum\yscale<\xscale\xscale=\yscale\fi
              \fi
   \fi
   \divide\drawingwd by1000 \multiply\drawingwd by\xscale
   \divide\drawinght by1000 \multiply\drawinght by\xscale
   \divide\psxoffset by1000 \multiply\psxoffset by\xscale
   \divide\psyoffset by1000 \multiply\psyoffset by\xscale
   \global\divide\pscm by 1000
   \global\multiply\pscm by\xscale
   \multiply\pswdincr by\xscale \multiply\pshtincr by\xscale
   \ifdim\d@mx=0pt\d@mx=\pswdincr\fi
   \ifdim\d@my=0pt\d@my=\pshtincr\fi
   \message{scaled \the\xscale}%
 \hbox to\d@mx{\hss\vbox to\d@my{\vss
   \global\setbox\drawingBox=\hbox to 0pt{\kern\psxoffset\vbox to 0pt{%
      \kern-\psyoffset
      \PSspeci@l{\PSfilename}{\the\xscale}%
      \vss}\hss\ps@nnotation}%
   \global\wd\drawingBox=\the\pswdincr
   \global\ht\drawingBox=\the\pshtincr
   \global\drawingwd=\pswdincr
   \global\drawinght=\pshtincr
   \baselineskip=0pt
   \copy\drawingBox
 \vss}\hss}%
  \global\psxoffset=0pt
  \global\psyoffset=0pt
  \global\pswdincr=0pt
  \global\pshtincr=0pt % These are local to one figure
  \global\pscm=1cm %should not be necessary
}}%
\def\psboxscaled#1#2{\vbox{%
  \ReadPSize{#2}%
  \xscale=#1
  \message{scaled \the\xscale}%
  \divide\pswdincr by 1000 \multiply\pswdincr by \xscale
  \divide\pshtincr by 1000 \multiply\pshtincr by \xscale
  \divide\psxoffset by1000 \multiply\psxoffset by\xscale
  \divide\psyoffset by1000 \multiply\psyoffset by\xscale
  \divide\drawingwd by1000 \multiply\drawingwd by\xscale
  \divide\drawinght by1000 \multiply\drawinght by\xscale
  \global\divide\pscm by 1000
  \global\multiply\pscm by\xscale
  \global\setbox\drawingBox=\hbox to 0pt{\kern\psxoffset\vbox to 0pt{%
     \kern-\psyoffset
     \PSspeci@l{\PSfilename}{\the\xscale}%
     \vss}\hss\ps@nnotation}%
  \advance\pswdincr by \drawingwd
  \advance\pshtincr by \drawinght
  \global\wd\drawingBox=\the\pswdincr
  \global\ht\drawingBox=\the\pshtincr
  \global\drawingwd=\pswdincr
  \global\drawinght=\pshtincr
  \baselineskip=0pt
  \copy\drawingBox
  \global\psxoffset=0pt
  \global\psyoffset=0pt
  \global\pswdincr=0pt
  \global\pshtincr=0pt % These are local to one figure
  \global\pscm=1cm
}}%
\def\psbox#1{\psboxscaled{1000}{#1}}%
\newif\ifn@teof\n@teoftrue
\newif\ifc@ntrolline
\newif\ifmatch
\newread\j@insplitin
\newwrite\j@insplitout
\newwrite\psbj@inaux
\immediate\openout\psbj@inaux=psbjoin.aux
\immediate\write\psbj@inaux{\string\joinfiles}%
\immediate\write\psbj@inaux{\jobname,}%
\def\toother#1{\ifcat\relax#1\else\expandafter%
  \toother@ux\meaning#1\endtoother@ux\fi}%
\def\toother@ux#1 #2#3\endtoother@ux{\def\tmp{#3}%
  \ifx\tmp\@mpty\def\tmp{#2}\let\next=\relax%
  \else\def\next{\toother@ux#2#3\endtoother@ux}\fi%
\next}%
\let\readfilenamehook=\relax
\def\re@d{\expandafter\re@daux}% spares typing 10 \expandafter's...
\def\re@daux{\futurelet\nextchar\stopre@dtest}%
\def\re@dnext{\xdef\lastreadfilename{\lastreadfilename\nextchar}%
  \afterassignment\re@d\let\nextchar}%
\def\stopre@d{\egroup\readfilenamehook}%
\def\stopre@dtest{%
  \ifcat\nextchar\relax\let\nextread\stopre@d
  \else
    \ifcat\nextchar\space\def\nextread{%
      \afterassignment\stopre@d\chardef\nextchar=`}%
    \else\let\nextread=\re@dnext
      \toother\nextchar
      \edef\nextchar{\tmp}%
    \fi
  \fi\nextread}%
\def\readfilename{\bgroup%
  \let\\=\backslashother \let\%=\percentother \let\~=\tildeother
  \let\#=\sharpother \xdef\lastreadfilename{}%
  \re@d}%
\xdef\GlobalInputList{\jobname}%
\def\psnewinput{%
  \def\readfilenamehook{% each entry in \GlobalInputList should be unique
    \if\matchexpin{\GlobalInputList}{, \lastreadfilename}%
    \else\xdef\GlobalInputList{\GlobalInputList, \lastreadfilename}%
      \immediate\write\psbj@inaux{\lastreadfilename,}%
    \fi%
    \ps@ldinput\lastreadfilename\relax%
    \let\readfilenamehook=\relax%
  }\readfilename%
}%
\expandafter\ifx\csname @@input\endcsname\relax    % then Plain
  \immediate\let\ps@ldinput=\input\def\input{\psnewinput}%
\else
  \immediate\let\ps@ldinput=\@@input
  \def\@@input{\psnewinput}%
\fi%
\def\nowarnopenout{%
 \def\warnopenout##1##2{%
   \readfilename##2\relax
   \message{\lastreadfilename}%
   \immediate\openout##1=\lastreadfilename\relax}}%
\def\warnopenout#1#2{%
 \readfilename#2\relax
 \def\t@mp{TrashMe,psbjoin.aux,psbjoint.tex,}\uncatcode\t@mp
 \if\matchexpin{\t@mp}{\lastreadfilename,}%
 \else
   \immediate\openin\pst@mpin=\lastreadfilename\relax
   \ifeof\pst@mpin
     \else
     \errhelp{If the content of this file is so precious to you, abort (ie
press x or e) and rename it before retrying.}%
     \errmessage{I'm just about to replace your file named \lastreadfilename}%
   \fi
   \immediate\closein\pst@mpin
 \fi
 \message{\lastreadfilename}%
 \immediate\openout#1=\lastreadfilename\relax}%
{\catcode`\%=12\catcode`\*=14
\gdef\splitfile#1{*
 \readfilename#1\relax
 \immediate\openin\j@insplitin=\lastreadfilename\relax
 \ifeof\j@insplitin
   \message{! I couldn't find and split \lastreadfilename!}*
 \else
   \immediate\openout\j@insplitout=TrashMe
   \message{< Splitting \lastreadfilename\space into}*
   \loop
     \ifeof\j@insplitin
       \immediate\closein\j@insplitin\n@teoffalse
     \else
       \n@teoftrue
       \executeinspecs{\global\read\j@insplitin to\spl@tinline\expandafter
         \ch@ckbeginnewfile\spl@tinline%Beginning-Of-File-Named:%\endcheck}*
       \ifc@ntrolline
       \else
         \toks0=\expandafter{\spl@tinline}*
         \immediate\write\j@insplitout{\the\toks0}*
       \fi
     \fi
   \ifn@teof\repeat
   \immediate\closeout\j@insplitout
 \fi\message{>}*
}*
\gdef\ch@ckbeginnewfile#1%Beginning-Of-File-Named:#2%#3\endcheck{*
 \def\t@mp{#1}*
 \ifx\@mpty\t@mp
   \def\t@mp{#3}*
   \ifx\@mpty\t@mp
     \global\c@ntrollinefalse
   \else
     \immediate\closeout\j@insplitout
     \warnopenout\j@insplitout{#2}*
     \global\c@ntrollinetrue
   \fi
 \else
   \global\c@ntrollinefalse
 \fi}*
\gdef\joinfiles#1\into#2{*
 \message{< Joining following files into}*
 \warnopenout\j@insplitout{#2}*
 \message{:}*
 {*
 \edef\w@##1{\immediate\write\j@insplitout{##1}}*
\w@{% This collection of files was produced with CERN psbox package}*
\w@{% To decompose and tex it:}*
\w@{%-save this with a filename CONTAINING ONLY LETTERS and a .TEX}*
\w@{% extension (say, JOINTFIL.TEX), in some uncrowded directory;}*
\w@{%-make sure you can \string\input\space psbox.tex (version>=1.3);}*
\w@{%  (else ftp cs.nyu.edu(=128.122.140.24):pub/TeX/psbox/, then get}*
\w@{%  and tex the file psboxall.tex; more info in psbREAD.ME)}*
\w@{%-tex JOINTFIL.TEX using Plain, or LaTeX, or whatever is needed by}*
\w@{%  the first file in the joining (after splitting JOINTFIL.TEX into}*
\w@{%  it's constituents, TeX will try to process it as it stands).}*
\w@{\string\input\space psbox.tex}*
\w@{\string\splitfile{\string\jobname}}*
\w@{\string\let\string\autojoin=\string\relax}*
}*
 \expandafter\tre@tfilelist#1, \endtre@t
 \immediate\closeout\j@insplitout
 \message{>}*
}*
\gdef\tre@tfilelist#1, #2\endtre@t{*
 \readfilename#1\relax
 \ifx\@mpty\lastreadfilename
 \else
   \immediate\openin\j@insplitin=\lastreadfilename\relax
   \ifeof\j@insplitin
     \errmessage{I couldn't find file \lastreadfilename}*
   \else
     \message{\lastreadfilename}*
     \immediate\write\j@insplitout{%Beginning-Of-File-Named:\lastreadfilename}*
     \executeinspecs{\global\read\j@insplitin to\oldj@ininline}*
     \loop
       \ifeof\j@insplitin\immediate\closein\j@insplitin\n@teoffalse
       \else\n@teoftrue
         \executeinspecs{\global\read\j@insplitin to\j@ininline}*
         \toks0=\expandafter{\oldj@ininline}*
         \let\oldj@ininline=\j@ininline
         \immediate\write\j@insplitout{\the\toks0}*
       \fi
     \ifn@teof
     \repeat
   \immediate\closein\j@insplitin
   \fi
   \tre@tfilelist#2, \endtre@t
 \fi}*
}%
\def\autojoin{%
 \immediate\write\psbj@inaux{\string\into{psbjoint.tex}}%
 \immediate\closeout\psbj@inaux
 \expandafter\joinfiles\GlobalInputList\into{psbjoint.tex}%
}%
\def\centinsert#1{\midinsert\line{\hss#1\hss}\endinsert}%
\def\psannotate#1#2{\vbox{%
  \def\ps@nnotation{#2\global\let\ps@nnotation=\relax}#1}}%
\def\pscaption#1#2{\vbox{%
   \setbox\drawingBox=#1
   \copy\drawingBox
   \vskip\baselineskip
   \vbox{\hsize=\wd\drawingBox\setbox0=\hbox{#2}%
     \ifdim\wd0>\hsize
       \noindent\unhbox0\tolerance=5000
    \else\centerline{\box0}%
    \fi
}}}%
\def\at(#1;#2)#3{\setbox0=\hbox{#3}\ht0=0pt\dp0=0pt
  \rlap{\kern#1\vbox to0pt{\kern-#2\box0\vss}}}%
\newdimen\gridht \newdimen\gridwd
\def\gridfill(#1;#2){%
  \setbox0=\hbox to 1\pscm
  {\vrule height1\pscm width.4pt\leaders\hrule\hfill}%
  \gridht=#1
  \divide\gridht by \ht0
  \multiply\gridht by \ht0
  \gridwd=#2
  \divide\gridwd by \wd0
  \multiply\gridwd by \wd0
  \advance \gridwd by \wd0
  \vbox to \gridht{\leaders\hbox to\gridwd{\leaders\box0\hfill}\vfill}}%
\def\fillinggrid{\at(0cm;0cm){\vbox{%
  \gridfill(\drawinght;\drawingwd)}}}%
\def\textleftof#1:{%
  \setbox1=#1
  \setbox0=\vbox\bgroup
    \advance\hsize by -\wd1 \advance\hsize by -2em}%
\def\textrightof#1:{%
  \setbox0=#1
  \setbox1=\vbox\bgroup
    \advance\hsize by -\wd0 \advance\hsize by -2em}%
\def\endtext{%
  \egroup
  \hbox to \hsize{\valign{\vfil##\vfil\cr%
\box0\cr%
\noalign{\hss}\box1\cr}}}%
\def\frameit#1#2#3{\hbox{\vrule width#1\vbox{%
  \hrule height#1\vskip#2\hbox{\hskip#2\vbox{#3}\hskip#2}%
        \vskip#2\hrule height#1}\vrule width#1}}%
\def\boxit#1{\frameit{0.4pt}{0pt}{#1}}%
\catcode`\@=12 % cs containing @ are unreachable
 \psfordvips   %     For DVIPS converter on VAX and UNIX

%% file: two.bbl
\begin{thebibliography}{21}
\bibitem[1]{CGM}{Charron E, Giusti-Suzor A and Mies F H 1993 {\it Phys. Rev.
Lett.}, {\bf 71}, 692.}                 
\bibitem[2]{YKW} {Yang X, Kim E K  and Wodtke A M  1990 {\it
J. Chem. Phys.} {\bf 93}, 4483.}
\bibitem[3]{SKSB}{Schiemann S, Kuhn A,
Steuerwald S and Bergmann K 1993, {\it
Phys. Rev. Lett.}, {\bf 71}, 3637.}
%\bibitem[4]{M}{Morse P.M. 1929 {\it Phys. Rev.} {\bf 34}, 57.}
\bibitem[4]{T}{Ting J J-L 
1994 {\it J. Phys. B} {\bf 27}, 1249.}
\bibitem[5]{DN} {Daboul J and Nieto M M 1994 {\it Phys. Lett. A}, in
press, hepth 9405154.}
\bibitem[6]{DPB} {Dauxois T, Peyrard M  and Bishop A R 1993 {\it Phys.
Rev. E} {\bf 47}, 684.}
\bibitem[7]{HM2} {Heather R   and Metiu H  1987 {\it J. Chem.
Phys.} {\bf 86}, 5009.}
%\bibitem[7]{HM1} {Heather R.  and Metiu H. 1988
% {\it J. Chem. Phys.} {\bf 88}, 5496.}
%\bibitem[8]{TM2}{Tanner J.J. and  Maricq M.M. 1988 {\it
%Chem. Phys. Lett. } {\bf 149},
%503.}
%\bibitem[9]{TM1}{Tanner J.J. and  Maricq M.M. 1989 {\it
%Phys. Rev. A } {\bf 40}, 4054.}
\bibitem[8]{TYJ}{Ting J J-L, Yuan J M,  and Jiang T F
1992 {\it Comp. Phys.  Comm.} {\bf 70}, 417.}
%\bibitem[11]{F}{Feit M.D., Fleck J.A. Jr. and
%Steiger A. 1982 {\it J. Comp. Phys.} {\bf 47}, 412.}
%\bibitem[13]{ALL}{Leforestier C. 1991 {\it et al},
%{\it J. Comp. Phys.} {\bf 94}, 59.}
%\bibitem[7]{GRSB}{Gaubatz U, Rudecki P, Schiemann S and Bergmann K 1990
%{\it J. Chem. Phys.}, {\bf 92}, 5363.}
%\bibitem[8]{MGHTW}{Melinger J S, Ganhdi S R, Hariharan A, Tull J X and 
%Warren W S 1992 {\it Phys. Rev.
%Lett.}, {\bf 68}, 2000.}
%\bibitem[8]{YW}{Yang X and Wodtke A M 1990 {\it
%J. Chem. Phys.}, {\bf 92}, 116.}
\bibitem[9]{RS}  {Schinke R 1993
{\it Photodissociation Dynamics}, Cambridge University Press, pp   371-374.}
\bibitem[10]{HH}{Huber K P   and Herzberg G  1979 {\it
Molecular Spectra and Molecular Structure
IV. Constants of Diatomic Molecules}, Van Nostrand Reinhold Co.}
\end{thebibliography}
